\documentclass[twocolumn]{article}
          \begin{document}
           \title{Language is Physical}
          \author{Paul Benioff\\
           Physics Division, Argonne National Laboratory \\
           Argonne, IL 60439 \\
           e-mail: pbenioff@anl.gov}
           \date{\today}
           \maketitle

          \begin{abstract}
          Some aspects of the physical nature of language are discussed. In particular,
          physical models of language must exist that are efficiently implementable.
          The existence requirement is essential because  without physical models
          no communication or thinking would be possible. Efficient implementability
          for creating and reading language expressions is discussed and illustrated
          with a quantum mechanical model. The reason for interest in language is that
          language expressions can have meaning, either as an informal language or as a
          formal language associated with mathematical or physical theories. It is noted
          that any universally applicable physical theory, or coherent theory of physics
          and mathematics together, includes in its domain physical models of expressions
          for both the informal language used to discuss the theory and the expressions of
          the theory itself. It follows that there must be some formulas in the formal theory
          that express some of their own physical properties.  The inclusion of intelligent
          systems in the domain of the theory means that the theory, e.g. quantum mechanics,
          must describe, in some sense, its own validation.  Maps of language expressions
          into physical states are discussed.  A spin projection example is discussed
          as are conditions under which such a map is a G\"{o}del map. The possibility
          that language is also mathematical is very briefly discussed.

          \end{abstract}

          %\pacs{03.67.-a,03.65.Ta,03.67.Lx}

          \section{Introduction}
          Quantum computation and quantum information are areas of much interest and
          research activity. The work began in 1980 -1982 \cite{Ben80} with the
         description of Hamiltonian  models of Turing machines in which the information
         bearing degrees of freedom only were described quantum mechanically.
         Since these models did not dissipate energy the work showed that the time energy
         uncertainty principle did not necessitate energy dissipation. Instead this
         principle relates the speed of a computation to the energy and energy  dispersion
         of the system being studied \cite{Margolus,Lloyd02}.

         Other work in the 1980s first suggested that computers operating quantum
         mechanically might be able to simulate physical systems much more efficiently
         than classical computers \cite{Feynman81}. In 1985 and 1989 the early models of
         quantum computation were expanded and generalized to the models based on
         networks of quantum gates and quantum Turing machines that are in wide use
         today as the basis for study of quantum computation and quantum information
         \cite{Feynman85,Deutsch}.

         An essential component of quantum computation and quantum information theory
         is the representation of numbers and information by quantum states. For
         example the state $|10010> =\otimes_{j=1}^{5}|s(j),j\rangle = |s\rangle$ where
         $j$ is the physical parameter denoting the place label and $s(1) =0,\cdots ,s(5)=1$
         is a binary representation of a number. Other representations of the number are
         $\sum_{j=1}^{5}2^{j-1}s(j)$ and $18$ in the decimal representation.  Linear
         superpositions of number states such as $\psi =\sum_{s}c_{s}|s\rangle$ where
         the sum is over all $2^{n}$ functions $s$ with domain $1,\cdots ,n$ and range
         $\{0,1\}$ play a very important role in quantum computation.

         Quantum information theory is based on the use of these states to also represent
         multiqubit states.  A qubit, which is the basic unit of quantum information
         theory, is represented by a binary state $\phi = \alpha|0\rangle +\beta|1\rangle$ where
         $|\alpha |^{2}+|\beta |^{2}=1$. Multiqubit states can be written as products
         of states of the form of $\phi$ or more generally as $\psi$ given above.

         A fundamental property of these qubit states is the fact that information is
         physical. This point, which was emphasized by Landauer \cite{Landauer}, states
         that all states representing information must have a physical basis. That
         is, information is necessarily represented by states of physical systems. An example
         often used is the representation of a qubit by the spin projection states of a
         spin $1/2$ system. If a physical representation were not possible, quantum
         information theory would not have any connection or relevance to physics.

         A similar situation holds for numbers, i.e. numbers are physical. This
         requirement means that numbers must be represented by
         states of physical systems.  This applies to natural numbers, integers, and
         rational numbers\footnote{Real and complex numbers, as limits of sequences of
         rational numbers, must be represented differently.}.

         One good way to characterize these different types of numbers is by the
         set of axioms for each type of numbers.  The natural numbers are described by
         the axioms of number theory or arithmetic, the integers and rational numbers
         by the respective axioms for a commutative ring with identity and a field
         (a ring with a multiplicative inverse) \cite{Adamson}.

         The requirement that numbers are physical means that there must exist physical
         systems with states such that the operations and relations defined by the
         axioms are physically implementable and that the physical implementations of
         the operations and relations  are efficient. This requirement of of efficient
         implementability  for an operation means that there must exist an actual
         physical procedure for carrying out the operation, and the space-time and
         thermodynamic resources required to carry out the procedure must be
         polynomial in the length of the numeral strings used to represent the
         numbers. In particular the resources required cannot be exponential in the
         string lengths.

         Physical representations and the requirement of efficient implementability for
         quantum systems have been discussed in more detail elsewhere for the natural
         numbers, integers, and rational numbers \cite{BenRNQM,BenRNQMALG}.
         There it was noted that the requirement of efficient implementability
         necessitates the use of $k-ary$ representations of numbers as the amount of
         information that can be contained and manipulated in a given space time
         volume is limited \cite{Lloyd1}.  Also, for arithmetic, the requirement of
         efficient implementability  necessitates the introduction of many successor
         operations, one for each $j$ corresponding to the addition of $k^{j-1}$,
         and axioms describing their properties. The reason is that the axioms of
         arithmetic describe just one successor, corresponding to addition of $1$, and
         the properties of the $+$ and $\times$ operations. The problem is that
         any procedure that adds or multiplies two numbers and is based on just the
         one successor operation is not efficient as it requires exponentially many
         iterations of the successor. In \cite{BenRNQMALG} these arguments
         were also seen to apply to representations of $+$ and $\times$ on states
         representing the integers and rational numbers.

         In this paper these arguments are extended to representations
         of languages in quantum mechanics.  That is, {\em language is physical}. This
         condition has two aspects: One is that physical models of language must exist
         and the other is that the physical models must be efficiently implementable.
         These are discussed in the next section where a specific model is
         summarized to illustrate efficient implementability.

         Probably the most important aspect of physical representations of language is
         that language can have meaning. This is the case if the language is  a formal
         language of a physical or mathematical theory, or is informal as a metalanguage
         used to discuss physical and mathematical theories in general. An example of
         the latter is the language used in this and other papers. Physical states,
         interpreted as expressions in an informal language or a formal language of a
         (consistent) theory, have meaning in that they describe properties of other
         systems (physical or mathematical) or even of themselves in a physical theory
         of sufficiently inclusive applicability.

         This is discussed more in Section \ref{PMT} where physical and mathematical
         theories in general are considered.  Included is a discussion of universal
         applicability of a physical theory and the need for a coherent theory of mathematics
         and physics  together that treats mathematics and physics together rather than as two
         entirely separate fields.

         The long term goal is to develop a coherent theory of physics and mathematics
         together.  At present this is just a dream \cite{Weinberg}. However there are
         various aspects of interest that can be said about this.  The above suggests
         the importance to such a theory, or to any universally applicable physical
         theory, of maps of language expressions into physical states.
         These are discussed in Section \ref{MLPS}. An example of such a map of
         expressions of any language into spin projection states of spin systems
         located on a 3D lattice is discussed.  The existence of many such maps and the
         conditions that such a map must satisfy if the map is to be a G\"{o}del map
         are discussed. For universally applicable physical theories it is noted that
         the necessary physical nature of language suggest that G\"{o}del maps may play
         a more limited role for these theories than for mathematical theories.

         A discussion of this is given in the last section. Included is the point that if
         language is mathematical, then conclusions similar to those reached for physical
         theories would apply to mathematical theories.  The observation that it seems
         unlikely that language is mathematical in the same immediate and necessary sense
         that it is physical is discussed. It is noted that one should keep an open mind on
         this question and that more work is needed.

         \section{Language is Physical} \label{LP}
         As noted in the introduction, the condition that language is physical means
         that physical models of language must exist and that the physical models must
         be efficiently implementable. That physical models must exist is quite evident.
         Without such models, communication would be impossible. Examples of such models
         include written text, speech, and optical transmission of language
         expressions (just as speech is transmission of language expressions by sound
         waves in air). Specific models of written text,  as distributions of
         ink molecules in a $3D$ lattice of potential wells, or spin projection states
         of systems in the same lattice, are discussed elsewhere
         \cite{BenTCTPM}.

         It is also quite likely that the ability to think or reason depends on the
         existence of physical models of language. Without entering into details of
         this complex subject it seems reasonable to expect that distinct conscious
         states of the brain correspond to distinct physical states of the brain. This
         would be expected to be the case independent of how one reasons or thinks
         (e.g. in picture sequences or word sequences, etc.). If such physical states
         did not exist, then it is likely that reasoning, thinking and even
         consciousness would not be possible \cite{Stapp,Squires,Page,Penrose}. That is,
         physical representations of language are a necessary, but probably not sufficient,
         condition for the existence of communication, thinking, and possibly even
         consciousness.

         It should be noted that here language is being given a very general definition
         in that it consists of  sets of words with the words in each set
         ordered into a sequence by their positions in space-time. Such an ordered
         sequence or words is referred to here as an expression. Individual words
         can take many forms, such  as strings of alphabet
         symbols, pictures, or signs (as in sign language for the deaf).

         The terminology used here is based on the set ${\mathcal A}^{*}$ of all finite
         strings of symbols in a finite alphabet $\mathcal A$. A particular symbol, $\#$, in
         $\mathcal A$ is singled out to represent a spacer symbol. An expression is any
         string in ${\mathcal A}^{*}$. An expression is a word  or formula if it does not
         contain $\#$. A string of words is an expression with words separated by one
         or more spacer symbols.

         Efficient implementability of physical models of a language means that
         physical states representing expressions (expression states) must be
         efficiently creatable and readable.  Also operations on the states that
         generate new expression states or linear superpositions of such  must be
         efficiently implementable.

         In general efficient implementability of a procedure means that there must
         exist a physically implementable dynamics for the procedure and the dynamics,
         represented by a unitary $U(t)=e^{-iHt}$, must be efficient.  That is, the
         space-time and thermodynamic resources needed to carry out the procedure must
         be polynomial in the expression length.  In particular it should not be
         exponential.

         An example to illustrate the meaning of efficient implementability consists of
         a multistate head $M$ moving on a $1D$ lattice of potential wells. Each well
         is occupied by one system with $k$ internal states corresponding to an
         alphabet $\mathcal A$ of $k$ symbols. Under the action of a unitary dynamics,
         $U_{M}(t)$, $M$ interacts locally with the systems and can move in either direction.

         Expression states have the form \begin{equation} |X,[a,b]\rangle =
         \otimes_{j=a}^{b}|X(j),j\rangle \label{expstate} \end{equation} where $X(j)\epsilon \mathcal A$
         and $[a,b]$ is an interval on the lattice. The Heisenberg representation
         projection operator for the state $|X,[a,b]\rangle$ at time $t$ is given by
         \begin{equation} P_{X,[a,b]}(t)=U_{M}^{\dagger}(t)P_{X,[a,b]}U_{M}(t).
         \end{equation}

         Consider an initial state $\Psi=|\underline{\#},i,0\rangle$
         consisting of all lattice systems in the spacer symbol $\#$ state and $M$ in
         state internal state $i$ and at position $0$.  Efficient creation of
         expression states means that for each $X,[a,b]$ the creation probability must
         be asymptotically stable. In other words, the limit \begin{equation}
         \langle \Psi |P_{X,[a,b]}(\infty)|\Psi\rangle = \lim_{t\rightarrow \infty}
         \langle \Psi |P_{X,[a,b]}(t)|\Psi\rangle\end{equation} must exist for each
         $X,[a,b]$.

         Efficient creation also means that the limit must be approached efficiently.
         The meaning of this is based on the definition of a limit: \begin{displaymath}
         \begin{array}{l}\forall m\exists \tau \forall
         t,t^{\prime}>\tau \\ \;\;| \langle \Psi |P_{X,[a,b]}(t)|\Psi\rangle -
         \langle \Psi |P_{X,[a,b]}(t^{\prime})|\Psi\rangle|<2^{-m}.\end{array} \end{displaymath}
         This states the existence, for each error bound, of a $\tau = \tau(X,m,a,n)$
         which is a lower bound for all times $t,t^{\prime}$ satisfying the inequality above. The
         dependence of $\tau$ on all the parameters, where $n$ is the length $L(X)$ of $X$, is
         indicated.

         Let $\tau(m,a,n) = \max_{X:L(X)=n}\tau(X,m,a,n)$.  Clearly the limit
         definition holds with $\tau(m,a,n)$ replacing $\tau(X,m,a,n)$. The requirement
         of efficient creation means that the $n$ dependence of $\tau(m,a,n)$ must
         satisfy \begin{equation} \tau(m,a,n)= K_{m,a}n^{\ell}.\label{taudep}
         \end{equation} In particular $\tau(m,a,n)$ must not have an exponential
         dependence on $n$ or $\tau(m,a,n) \neq C _{m,a}2^{n^{\nu}}.$
         Here the $m$ and $a$ dependence is included in the
         constants $K_{m.a},C_{m,a}$.  Typically the polynomial and exponential
         exponents $\ell\sim 1-3$ and $\nu\sim 1$.  Factors of $\log
         n$ have been ignored in the above.

         This states explicitly that for each $m$, the time needed for the probability
         of creating an expression at a given lattice point to be within $2^{-m}$ of a
         limit probability, has a polynomial dependence on the length of the expression.
         That is, the time resources needed to create an expression at a given lattice
         point with some probability $p$ have a polynomial dependence on the length of
         the expression.

         Note that this says nothing about the dependence of the time resources needed
         on either the accuracy or value of $m$ or on the lattice position $a$.  For example
         it could be that $K_{m,a}$ in Eq. \ref{taudep} has an exponential dependence
         on $m$ or $K_{m,a}=k_{a}2^{m}$.

         The above example refers to efficient creation of an expression state.  The
         same considerations apply to any operation that involves reading an expression
         state.  In particular reading rules for expressions that are used are very
         simple and minimize the energy and momentum resources required to read an expression.
         An example of such a rule is the motion of a head $M$ along a straight line
         path in a  $2D$ or $3D$ lattice to read an expression state $|X,[a,b]\rangle$.
         Carrying this out expends energy and momentum at a rate that is polynomial in
         the length $n$ of $X$ where the polynomial exponent is $\sim 1$.

         The point of this is to emphasize that such simple reading rules, which are
         universally used, are the exception in that almost all reading rules are much
         more complex and require much more space time and thermodynamic resources to
         implement. To see this consider that an expression state $|X,[a,b]\rangle =
         \otimes_{j=a}^{b}|X(j),j\rangle$ is a special case of a state
         $|X,p\rangle=\otimes_{j=1}^{n}|X(j),p(j)\rangle$ where $p$ is a function from
         the positive integers to the points on the lattice and $n$ is the length of
         $X$. Clearly any reading rule based on a general $p$ has $M$ moving on the lattice in
         many different directions, from $p(j)$ to $p(j+1)$ for $j=1,2,\cdots$, and
         expending much more energy and momentum in the process.

         The reading rule can be even more complex in that the location of the lattice
         chosen for the $j+1st$ symbol of $X$ can depend on the first $j$ symbols
         read in $X$. In this case
         $|X,p\rangle=\otimes_{j=1}^{n}|X(j),p(X_{[1,j-1]},j)\rangle$.  Here
         $X_{[1,j-1]}$ denotes the first $j-1$ symbols of $X$ and $p $ is a function from
         $N\times {\mathcal A}^{*}$ to the lattice. In this case it would be
         exponentially hard for a head $M$ to use rule $p$ to read $|X,p\rangle$. These
         arguments show clearly why simple reading rules are universally used
         even though these rules are in the minority as most rules are not simple.

         So far the discussion about efficient implementability is not unique to
         language as it applies also to $k-ary$ representations of numbers.  The
         interest in language representations is based on the observation that language
         expressions can have meaning. This is the case for language expressions as formulas,
         words, and word strings in physical and mathematical theories and as expressions in
         the  informal language used to describe theories and other aspects of existence.

         This aspect is the basis for interest in the condition that language is
         physical, and in physical and mathematical theories in general.  Some relevant aspects
         of these theories are discussed in the next section.

         \section{Physical and Mathematical Theories} \label{PMT}
         Here the interest is in theories that are first order axiomatizable. These are
         theories based on axiom sets for which the range of variables is limited to
         individuals in a model domain It does not also include sets of individuals as in
         second order logic \cite{Vaananen}. The limitation to first
         order theories is made because most studies in mathematical logic are
         concerned with first order theories.

         Mathematical logic is the study of various axiomatizable mathematical theories and
         the relationship between such concepts as truth, validity, consistency,
         completeness and provability. Here the interest is in extending these concepts
         to physical theories. (See \cite{BenUMLCQM} for a very simple example of such
         an extension.) For the purposes of this paper the details of specific sets of
         axioms are not relevant.

         Here interest is in those language expression states, such as $|X,[a,b]\rangle$
         that are formula states (or word states), term states, or expression states
         representing strings of formulas in languages of physical and mathematical theories.
         In this case, by G\"{o}del's completeness theorem \cite{Shoenfield}, the formula
         states have meaning if and only if the theory being considered is consistent.

         In this case the formulas\footnote{From now the explicit reference to "states"
         for states representing formulas, terms, etc., will be suppressed.} acquire meaning
         through an interpretation into a  domain of applicability of the
         theory. An interpretation $I$ is a map of the symbols (or symbol strings if needed)
         for constants, variables, operations or functions, and relations of the theory
         language into elements, variables, basic operations or functions,
         and basic relations in a structure. In this way variable free terms are mapped
         into elements, terms with variables are mapped into functions, and  formulas are
         mapped into relations in the structure.  The structure or
         domain of applicability of the theory is a model of the theory if all the
         axioms of the theory are true for the map $I$. For a given theory there are
         many different models each with its associated map $I$.  Some of these models
         are isomorphic to one another and (for first order theories) others are not.
         Additional details are given in \cite{Shoenfield}.

         Here the interest is in physical theories that are universally applicable.
         Quantum mechanics, or some suitable generalization such as quantum field
         theory, is an example of such a theory. It is not known at present how to
         define universal applicability or even if this concept is useful.  One
         problem with a simple definition that states that the theory is applicable to all
         physical systems, with no restrictions on what one means by "all", is that,
         according to Tegmark \cite{Tegmark}, it leads to acceptance of the Everett
         Wheeler interpretation of quantum mechanics. This follows logically from the two
         statements: "All isolated systems are described by the Schr\"{o}dinger equation in
         quantum mechanics", " The universe is an isolated system" \cite{Tegmark}.

         For the purposes of this paper it is sufficient to require that the domain of
         applicability include intelligent beings as systems occupying a finite region
         of space-time.  Also included are physical systems with states that
         can represent language.  That is, the states must be such that creating and
         reading expression states, and other operations on expression states, must be
         efficiently implementable. The same requirement holds for numbers in that all
         physical systems with states that can represent numbers of different types
         must be included. From here on it is assumed that however one defines
         universal applicability, it must satisfy these requirements.

         The requirements of universal applicability of a physical theory and language
         is physical have the consequence that there exist physical systems with states
         representing language expressions, and that these systems are in a model domain
         of the theory. This applies to both the informal language used by an intelligent being
         to describe such a theory, (or any other theory) and to the formulas and theorems
         of the formal language of the theory.  For any representation of the
         informal language  and formal language of the theory, there must be formulas and
         theorems of the theory that describe the physical properties of  the representations
         of expressions of both the informal and formal languages. In particular there must
         be some formulas and theorems of the theory that describe some of their own
         physical properties.

         Note that the formulas and theorems with this self description property are
         different for different physical representations of the language expressions.
         This follows from the observation that representing expressions or text as
         arrangements of ink molecules on a 3D lattice of potential wells (text on
         printed pages) has a different physical description than representation as
         modulated sound waves (speech).

         Since quantum mechanics is supposed to be universally applicable, the
         expressions, formulas and theorems of any theory are represented by states of
         quantum systems.  This applies to quantum mechanics itself as a theory in that
         some states of physical systems can be interpreted as describing their own
         physics.

         It also follows from the universal applicability of quantum mechanics that
         intelligent systems are quantum systems.  This is the case even
         if they are macroscopic systems of about $10^{25}$ degrees of freedom.

         An important activity of intelligent quantum systems is the construction of a
         valid physical theory using physics and mathematics. If the theory is
         universally applicable then the theory being validated is the theory
         describing the dynamics of the systems carrying out the validation.  The
         theory must in some sense describe its own validation.  In particular quantum
         mechanics must, in some sense, describe its own validation.

         The details of this description and what is involved are not known at present.
         However, as is well known, an important part of the validation process is the
         comparison of theoretical predictions with experimental results. The
         theoretical predictions, which are based on mathematical theorems and formulas
         in the mathematics of quantum mechanics, are determined by computations on a
         computer.  It follows from the universality of quantum mechanics that
         computations are a dynamical quantum process on a quantum system, and that
         this process should be described within quantum mechanics \footnote{It is
         this group of ideas that led to the authors early work on quantum mechanical
         models of computation.}.

         Similarly the experimental process, consisting of preparation of a system in
         some state and measuring some property of it, is a quantum dynamical process.
         Both the apparatus used to prepare a system in some state (if preparation is
         part of the measurement) and the apparatus used to measure some property of
         the system so prepared, are quantum systems. It follows that the dynamics of
         the preparation and measurement procedures are quantum dynamical procedures
         that should be describable within quantum mechanics just as the computation
         process is describable within quantum mechanics.

         It must be emphasized that it follows from the universality of quantum
         mechanics that measurements and computational processes are a small  fraction
         of processes in general whose dynamics are described in quantum mechanics.
         Most processes are meaningless and are neither experiments nor computations.
         And quantum mechanics itself seems to be completely silent on which processes have
         meaning as experiments or as computations, or are meaningless.

         An essential role provided by  intelligent quantum systems is to determine
         which processes have meaning as experiments and which have meaning as
         computations. These systems determine which physical quantum processes are
         valid computations and which are valid experiments. As part of the validation
         process for quantum mechanics, intelligent systems
         also determine which computation should be compared with which experiment.
         Since intelligent systems are quantum systems, a goal of future work is to
         describe, to the maximum extent possible, the validation process itself
         within quantum mechanics.

         As part of such a description it is to be noted that the validation process
         depends on the close intertwining of physics and mathematics.  This suggests a
         need for a {\em coherent theory of mathematics and physics together} that
         describes, to the maximum extent possible, its own validity and strength and
         is maximally valid and strong. The universal applicability of quantum
         mechanics suggests that development of a coherent theory of mathematics and
         quantum mechanics together is needed.  Such a theory should include, to the
         maximum extent possible, a quantum dynamical description of its own validation
         by quantum systems.  It should also maximally describe its own validity and
         strength and be maximally valid and strong. (See \cite{BenTCTPM}for more
         discussion of these ideas.)

        \section{Maps of Language into Physical States}\label{MLPS}
        It is expected that an important component of a coherent theory
        of physics and mathematics will consist of maps from language expressions to
        states of physical systems. To this end some very simple aspects of maps from
        language expressions to states of physical systems will be examined in this section.
        In particular the interest is in maps from the expressions  of any language, formal or
        informal, to states of quantum systems. The states can be macroscopic, as in the case
        of computers used so far, or they can be microscopic, as in models
        of quantum computation.

         A simple example of a map $G$ to microscopic physical states can be described for
         quantum spin systems of spin $s$ located on a $3D$ lattice $\mathcal L$ of space points $
         \underline{x}=x,y,z$. The spin $s$ is such that $2s+1\geq |{\mathcal A}|$, the
         number of symbols in the language alphabet $\mathcal A$. The map $G$ from expressions
         can be defined from two $1-1$ maps $g:{\mathcal A}\rightarrow \{|m\rangle:
         -s\leq m \leq s\}$ and $p:N\rightarrow \{|\underline{x}\rangle :
         \underline{x}\epsilon {\mathcal L}\}$.  Here $g$ maps the alphabet to single system
         spin projection states, $p$ is a map from the natural numbers to space point
         states on the lattice $\mathcal L$, and ${\mathcal A}^{*}$ is the set of all
         expressions as finite length strings of symbols in $\mathcal A$.

         The map $G$ is defined by the pair $(g,p)$ such that \begin{equation} G(X)=
         \otimes_{j=1}^{|X|}|g(X(j)),p(j)\rangle. \label{exgodel}\end{equation} Here
         $X$, as a symbol string, is a function from $1,2,\cdots, |X|$ to $\mathcal A$,
         $|X|$ is the number of symbols in $X$, and $|g(X(j)),p(j)\rangle
         =|g(X(j)\rangle\times|p(j)\rangle$ is the state of a spin system at site
         $p(j)$. $p$ represents the path in $\mathcal L$ along which the spin systems are
         located.  The ordering of the symbols in $X$ into a symbol string and the
         ordering of the spin states along $p$ is based on the canonical well ordering of
         the natural numbers in $N$.

         Several features of $G$ should be noted.  There are a great many maps $G$
         from the expressions in ${\mathcal A}^{*}$ to states of quantum mechanical
         systems. Besides the existence of many different maps $p$, the map $g$ can change
         in many ways. For example, let $u$ be any unitary operator on the $|{\mathcal A}|$
         dimensional Hilbert space of each spin system and define $g_{u} = ug$. This
         defines a new map $G_{u}$ according to $G_{u}(X)=
         \otimes_{j=1}^{|X|}|u(g(X(j))),p(j)\rangle$.

         Other possibilities for $g$ include maps of alphabet symbols into
         entangled spin states of several quantum  spin systems.  This type of map is useful in
         quantum error correction schemes \cite{QEC}. One can also let $u=u_{
         \underline{x}}$ depend on the lattice locations.  This type of map is similar to the
         local gauge transformations used in quantum field theory.

         Also $g$ can map the alphabet to quite different types of microscopic states of
         quantum systems.  $g$ can also map the alphabet to macroscopic states of
         systems. An example describes symbols $A$ in $\mathcal A$ as particular arrangements
         of ink molecules in a lattice of potential wells, one molecule per well. To each
         alphabet symbol $A$ is associated a density operator state $\rho_{A}$ characterized
         by a lattice location parameter and scale parameters to determine the size and shape
         of $A$. Additional details are given elsewhere \cite{BenTCTPM}.

         The above discussion might lead one to believe that any map $G$ from language
         expressions to physical states is acceptable. This is not the case as $G$
         does have some restrictions. It should be such that the syntactic operations,
         including creating and reading the states $G(X)$ (denoted as $|X,[1,|X|]\rangle$),
         are efficiently implementable.  This point has already been discussed in
         Section \ref{LP}.

         The definition of $G$ in Eq. \ref{exgodel} is independent of which expressions in
         ${\mathcal A}^{*}$ correspond to words, formulas, or theorems in a theory.
         It applies to any language of any theory based on the alphabet $\mathcal A$. The
         language can be that of a consistent or an inconsistent
         theory, or it can be informal such as the language used to write this paper.

         Note that the definition of $G$ maps expressions directly into physical system
         states without first mapping expressions into multiqunit states and then
         mapping multiqunit states into states of physical systems. The advantage of
         proceeding through states of qunits as units of quantum information with $n$ basis
         states is that qunits provide a good quantum mechanical reference point.  One can
         then consider mappings from qunit states to states of different physical
         systems. This is what is done in quantum computation work where the expressions are
         strings of numerals, usually in the binary basis.  Here this two step mapping
         is not used as it is not needed.  It is more economical to proceed directly
         from expressions in ${\mathcal A}^{*}$ to states of physical systems.

         As defined $G$ is a map from the expressions of any language into the physical states
         of quantum systems. It is of interest to restrict $G$ to the case that the
         language is associated with a  consistent theory and the physical states are
         part of a model universe or domain of applicability for the theory.
         Then $G$ maps expressions corresponding to terms, formulas, and strings of
         formulas to states of quantum systems in a domain or model of the theory.  In this
         case $G$ is an example of a G\"{o}del map.  These maps can be used to interpret
         formulas of the language as describing some of their own properties.

         In general a G\"{o}del map $G$ is a $1-1$ map of the expressions of a theory language
         into individual elements of any model for the theory.  By use of such a map formulas
         that describe properties of the elements of the model under $I$ can be interpreted
         through $G$ to describe syntactic
         properties of the language expressions. The canonical example, which was first
         used by G\"{o}del to prove incompleteness theorems for arithmetic \cite{Godel}, is
         a map from the expressions of the language of arithmetic to the set of natural
         numbers.  In this way arithmetic formulas, which describe properties of
         numbers, can be interpreted to describe properties of the language expressions
         themselves \cite{Smullyan,Shoenfield}.

         The concept of a G\"{o}del map can be extended to any physical or mathematical
         theory, not just arithmetic. For quantum mechanics a map $G$ would be a map from
         expressions in the language of quantum mechanics to states of quantum systems that
         are included in a model of a theory of quantum mechanics. In this way some formulas
         in the language of quantum mechanics that describe properties of quantum
         systems can be interpreted through $G$ to describe some of their own
         properties.

         It is not necessary that the individual model elements be states of quantum
         systems.  They can be other quantum objects such as operators or modes of
         quantum fields.  These may be useful for constructing quantum mechanical
         models of the theories or real and complex numbers.

        Additional aspects of these maps and their relation to interpretations of
        theories will be left to future work. Here it is sufficient to note that for a
        universally applicable theory such as quantum mechanics, it follows from the
        observation that language is physical that any expression of any language for
        quantum mechanics is already a physical state of systems in the domain of the
        theory.  In this case one would expect G\"{o}del maps to have a different
        character than for the case where the theory is not about quantum systems.  An
        example of this would be the case where the theory is arithmetic
        which is about numbers and not about quantum systems. More details about these
        aspects will be addressed in future work.

        \section{Discussion}\label{D}

         It is of interest to ask if G\"{o}del maps provide the same
         function for mathematical theories as for universally applicable physical
         theories. That language is physical suggests that there may be a difference.
         For universally applicable physical theories, all physical representations of
         language are necessarily already in the model domain of the theory. For any
         representation there are formulas that describe physical properties of the
         systems in the representation. In this case the role of $G$ is limited to
         determining the particular correspondence $X\rightarrow G(X)$. The range set of
         $G$ is already included in the space of states of physical properties of systems
         in the model domain.  Eq. \ref{exgodel} is an example in which the space of
         physical properties is the Hilbert space spanned by the spin projection eigenstates
         of the lattice spin systems.

         For  mathematics the situation equivalent to that in physics would be the existence
         of a universally applicable mathematical theory and the condition that language is
         mathematical.  That is, there must necessarily exist many mathematical representations
         of the language all of which are contained in a model of a universally applicable
         theory. It is not at all clear that this is the case.

         Zermelo Frankel set theory is an example of a very powerful mathematical
         theory.Even though it does not include all of mathematics \cite{Frankel} its
         domain of applicability is sufficiently broad to be considered universally
         applicable from the viewpoint taken here. Also it includes all the mathematics
         used so far by physics \cite{BenJMP}. However it is not at all clear that
         languages must have mathematical representations in an immediate and
         direct sense that it must have physical
         representations. As noted before in Section \ref{LP}, The fact that we can
         communicate, e.g. by speech or writing, and can think  implies the existence
         of physical representations of language. If these representations did not exist
         we could not communicate or even think.

         From a mathematical point of view the existence or nonexistence of such
         representations is an abstract issue that has nothing to do with communication
         or any other basic physical or mathematical aspect. Also mathematical
         representations can always be constructed by use of G\"{o}del maps. However
         this does not address the question whether the potential nonexistence of
         mathematical representations of language has any immediate and direct
         consequences as is the case for physical representations.

         At this point it seems unlikely that language is mathematical in a same or similar
         sense that language is physical. However this is an open question. Further elucidation
         may have to await more development of a coherent theory of mathematics and physics
         together.

         If one regards language expressions as strings of symbols in an alphabet
         $\mathcal A$ as an abstract concept in that expressions have some ideal existence
         outside of space and time, then the maps $G$ link these abstract objects to
         states of physical systems.  This is similar to the realist position in
         mathematics that regards mathematical objects as having an ideal or
         Platonic existence outside of space and time. This position, which as been
         much discussed in the literature \cite{Frankel,MarMyc,Shapiro}, has its own
         set of problems.  It is clear that more work needs to be done regarding the
         role played by the maps $G$ linking language expressions to states of physical
         systems.

         \section*{Acknowledgements}
        This work is supported by the U.S. Department of Energy, Nuclear Physics Division,
        under contract W-31-109-ENG-38.

          \end{document}